# A Lightweight and Flexible Mobile Agent Platform Tailored to Management Applications

Dr. Damianos Gavalas

Department of Cultural Technology and Communication,
University of the Aegean,
Mytilene, Lesvos Island, Greece

E-mail: dgavalas@aegean.gr

**Abstract -** Mobile Agents (MAs) represent a distributed computing technology that promises to address the scalability problems of centralized network management. A critical issue that will affect the wider adoption of MA paradigm in management applications is the development of MA Platforms (MAPs) expressly oriented to distributed management. However, most of available platforms impose considerable burden on network and system resources and also lack of essential functionality, such as security mechanisms, fault tolerance, strategies for building network-aware MA itineraries and support for user-friendly customization of MA-based management tasks. In this paper, we discuss the design considerations and implementation details of a complete MAP research prototype that sufficiently addresses all the aforementioned issues. Our MAP has been implemented in Java and tailored for network and systems management applications.

## I. INTRODUCTION

Current IP network management systems are typically based on centralized client-server architectures, with the Simple Network Management Protocol (SNMP) [15] being the core management protocol. Within SNMP, the *manager* application serves the role of the client and the static SNMP *agents* (installed on managed devices) serve the role of distributed servers. The functionality of both the manager and the SNMP agents is rigidly defined at design time. Physical resources are represented by managed objects. Collections of managed objects are grouped into tree-structured Management Information Bases (MIB). Despite its wide deployment base, SNMP presents serious scalability problems when the size and complexity of the network increases [9][16]. Thus, the need for decentralized and distributed network management architectures is more important and necessary than ever before [17].

The emerging Mobile Agent (MA) technology can play an important role in distributed network and systems management (NSM) [5]. The term MA refers to a software unit that travels between network nodes following either a pre-defined or a context-dependent itinerary [6]. Post its creation, an MA can carry its *persistent state* and *code* to another node, where its execution can be restarted or resumed. Through interacting with a node, an agent can perform complex processing and filtering operations upon retrieved data, directly control equipment and dynamically deploy software to the nodes. That is, the agent can carry the application logic where it is needed and only accumulate filtered data rather than the entire data set retrieved from nodes [17]. Several researchers have proposed the application of MA technology in the area of NSM [3][4][9][11][13][14][16]. Typical scenarios involve monitoring multi-hop MAs that sequentially visit a set of nodes to retrieve and filter management data and then deliver high-level results to the manager host.

The phenomenal popularity of MAs is reflected on several industrial initiatives that led to the development of numerous Mobile Agent Platforms (MAPs) [5]. Most of them represent commercial, general-purpose MAPs, e.g. IBM Aglets [1] and Toshiba Bee-gent [2]. Among other areas, these platforms have been employed in distributed management applications [4][16]. From the management viewpoint though, general-purpose MAPs share several weaknesses: (a) they incorporate rich functionality, yet, usually unnecessary for management applications, thereby seriously affecting the usage of system and network resources; (b) some essential features (for instance, the ability of servers that receive incoming MAs to distinguish between different versions of the same MA class, which may reflect the update/modification of an existing management task) are not supported; (c) companies shipping commercial MAPs, often suspend their support.

The aforementioned weaknesses of general-purpose MAPs led to the development of numerous MAP research prototypes with network management orientation, aiming at optimizing functionality, flexibility and performance aspects, e.g. Codeshell [3], MAP [11] and MobileSpaces [14]. However, a number of limitations have been identified on these MAPs:

- *Heavyweight migration schemes*: Most existing MAPs involve the transfer of both state *and* code at each MA migration. However, the transfer of code is necessary only when the MA visits a device for a first time. That inefficient scheme may result in serious scalability problems both in terms of latency and migration overhead.
- *MA services customization*: The development and customization of MA-enabled management tasks is not effortless with available MAPs, as it requires programming skills and detailed knowledge of the MAPs' design.

- *Class loading*: Most MAPs include a class loader component, able to receive and load at runtime visiting MAs bytecode. Yet, to the best of our knowledge, there is not any MAP which allows to overwrite a cached MA class, i.e. to dynamically upgrade MA-enabled management tasks.
- *Security*: Not all management-oriented MAPs sufficiently address security issues related to MAs, e.g. authentication, authorization and encryption.
- *Lack of itinerary optimization* strategy: When multi-hop MA-based management tasks are concerned, the order in which MAs visit managed devices (i.e. the MAs' itinerary) is a crucial factor seriously affecting the overall delay and the network overhead imposed by the MA transfers [6][8]. Hence, an efficient MA itinerary design approach is needed, which is not addressed by any of the available MAPs.

In this paper, we introduce a Java-based MAP research prototype that addresses all the aforementioned limitations of existing general-purpose and management-oriented MAPs. The main strengths of our MAP are:

- Network management applications orientation: Unlike publicly available general-purpose MAPs, the introduced MAP takes into account the special characteristics and requirements of management applications and therefore provides a flexible and scalable environment tailored to distributed management operations.
- Lightweight memory and storage footprint (only the necessary MAP functionality for management operations is included).
- Support for existing management standards: Our architecture encompasses the dominant management framework of the Internet world, i.e. the SNMP. Due to its huge installation base, integration with SNMP was considered of vital importance to maintain compliance with legacy management systems [10].
- Adoption of modular MAP architecture that eases the addition of new services or the modification of existing ones.
- Implementation of a novel lightweight migration scheme wherein the MA code is deployed at startup and only the MA state transferred thereafter, resulting in minimal usage of network resources (bytecode size is typically much larger than state size [6]) and faster class loading.
- Incorporation of various migration optimization (programming) techniques suggested in [7].
- Ability of managed devices to distinguish between different versions of the same MA class, which may reflect the update/modification of an existing management task; to achieve that, a novel class-loading mechanism has been implemented as an extension to the standard Java Virtual Machine (JVM).
- Easy addition of new MA-based management tasks through a novel tool (mobile agent generator) which removes the requirement for programming skills by the network administrator.
- Support for essential security aspects (authentication, authorization and encryption).
- Implementation of a software module that, depending on the underlying network topology and the set of devices involved in a management task, constructs near-optimal MA itineraries thereby optimizing the usage of network resources.
- Our MAP has been entirely developed in Java to guarantee support for any network device, regardless of the underlying hardware platform or operating system.

The remainder of the paper is organized as follows: Section II comprises the core of this article, describing in detail the software components of the proposed MAP. Section III concludes the paper and draws direction for future research on the field.

## II. INFRASTRUCTURE OVERVIEW - IMPLEMENTATION DETAILS

Our MAP prototype consists of the following major building blocks (see Figure 1):

- The Manager, responsible for launching and controlling MAs and displaying results;
- The MA objects, capable of migrating between the managed entities to collect information based on pre-defined policies;
- The Mobile Agent Server (MAS), capable of receiving MAs and providing an interface to the local physical resources;
- The Mobile Agent Generator (MAG), a tool that automates the creation and deployment of service-oriented MA objects.

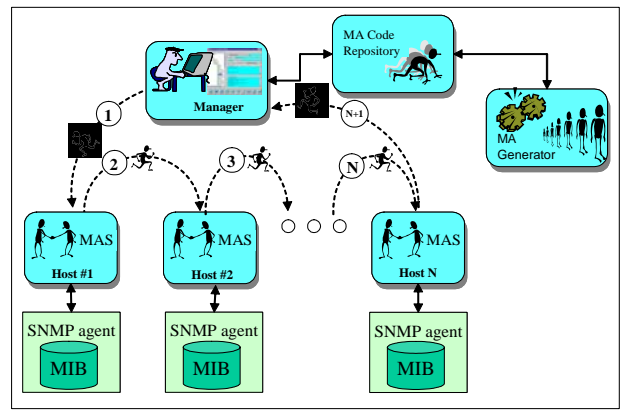

Figure 1. The Mobile Agents-based Infrastructure

The following sections elaborate on the design and implementation details behind its one of these components.

### A. Manager application

The manager application performs monitoring and control operations through interacting with devices running agent processes. It comprises a multithreading environment where a main thread instantiates, controls and co-ordinates the operation of a number of specialized threads (see Figure 2).

At startup, the manager's Network Discovery Thread executes a topology discovery algorithm to populate a list of 'discovered' active MASs. In the event of a pending MA-based task, the following steps are followed by the manager application:
- the class definition of the MA corresponding to the management task is retrieved from the Mobile Code Repository (MCR);
- the Itinerary Scheduler Module (ISM), which executes an algorithm that suggests the number *N* of MAs should be created to carry out this task and what their itinerary should be;
- a Polling Thread (PT) is instantiated, which is responsible for the execution of this particular management task;
- the PT instantiates *N* MA objects and assigns them their itinerary (provided by the ISM);
- the MAs' state information is then compressed and transferred through the Migration Facility Component (MFC) to their first destination host;
- when the MAs return back, they are received by the Mobile Agent Listener (MAL) thread, which retrieves the results carried by the MAs, presents them to the user and optionally stores them in a MySQL database.

The MAL, MFC and Security Component (SC) are identical to the ones used by the MAS entities; their implementation details are discussed in subsection C.

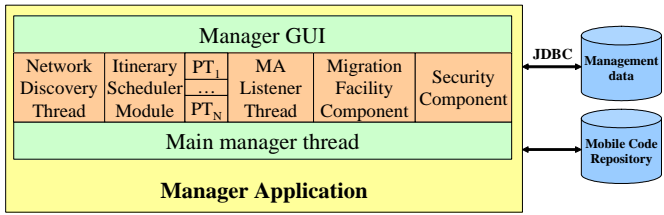

Figure 2. Break-down of the components that compose the manager application

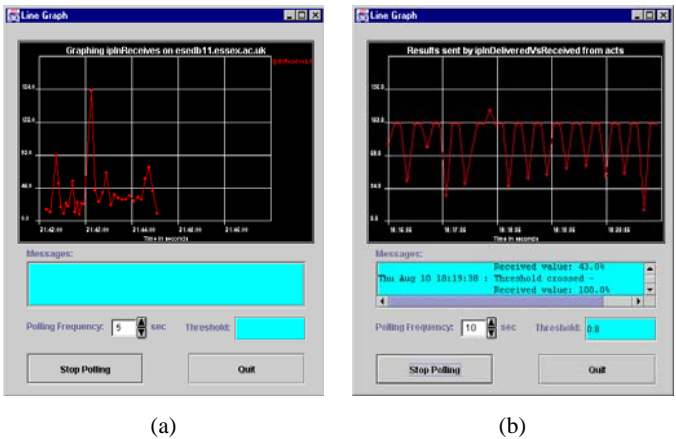

Figure 3. (a) Polling of a scalar MIB variable; (b) real-time notifications for management parameters crossing pre-determined thresholds.

The results delivered by monitoring MAs may be displayed in real-time line graphs. Figure 3 presents screenshots of real-time graphs displaying the fluctuation of a MIB scalar variable value retrieved from a specific managed device and notification messages (alarms) issued upon the event of management parameters crossing pre-determined thresholds.

**Itinerary Scheduler Module (ISM)**

In network monitoring applications, using a single MA object launched from the manager platform that sequentially visits the entire managed devices set, may result in serious scalability problems: the network overhead imposed by the MA transfers grows exponentially with the network size, since MAs accumulate more data from every visited node and become heavier on the next hop [6]. Hence, a more efficient MA itinerary design approach is needed, which should take into account the underlying managed network topology and the volume of accumulated management data; in such approach, multiple MAs (each assigned a limited-hop itinerary) will be employed in parallel to carry out given management tasks, thereby minimizing the overall response time and optimizing the use of network resources.

To this end, we have implemented the Heuristic algorithm for Itinerary Planning (HIP), which has been introduced and evaluated in [8]. HIP's Java implementation has been incorporated into the manager's Itinerary Scheduler Module (ISM). HIP is executed whenever a new multi-hop agent-based management task is about to start, suggesting not only the optimum number of MAs that should be dispatched but also their exact itinerary. It is noted that HIP's output is re-calculated whenever the managed network topology is altered, e.g. upon the event of a node's failure or the discovery of a new active agent server.

*B. Mobile Agent Implementation*

An MA object is identified by its code (description of its behaviour) and persistent state information (modifiable variables). In the context of our MAP, MAs are Java classes supplied with:
- a unique ID;
- an 'itinerary folder', including a list of hosts to be visited during the MA's operational travel; the itinerary is supplied by the ISM;
- a 'data folder', used to store collected data;
- a flag indicating whether retrieved data should be encrypted;
- a byte array containing the MA's 'signature', used for authenticating the MA when it arrives at a destination host;
- a number of methods that facilitate the interaction with managed devices.

Following the suggestions reported in [7], we have implemented a series of design optimizations aiming at minimizing MA migrations overhead and latency: (a) use of a lightweight MA code transfer scheme (described in subsection D), (b) reduced state size through minimal use of non-transient objects and use of primitive instead of complex data types for non-transient objects (e.g. use of arrays instead of Vector objects), (c) compression of MA state

(using the Java gzip utility) and use of stream buffering for MA transfers.

MA transfers are performed by the MFC components. In particular, through the process of *serialization*, the state of an MA object can be saved, transferred through the network and reconstructed (de-serialized) at the receiving node. To protect MAs against tampering, sensitive MA properties may be specified only once, when the MA is created. If a malicious host attempts to modify these properties, a Not_Authorized_To_Initialize_Exception exception is thrown. There are also several callback methods invoked when the MA object is instantiated, arrives or moves from a host, fails to migrate, or when its execution is started, stopped, resumed or suspended.

*C. The Mobile Agent Server (MAS)*

The interface between visiting MAs and legacy systems is implemented through MAS modules[1]. Functionally, the MASs reside above standard SNMP agents, defining an efficient runtime environment for receiving, instantiating, executing, and dispatching incoming MA objects, whilst protecting the system against malicious agent attacks. The SNMP agent process is started automatically at MAS initialization, if not already active.

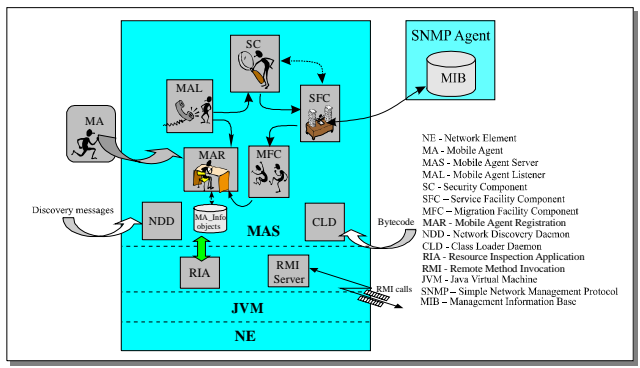

Figure 4. The Mobile Agent Server

Similarly to the manager application, the MAS class is designed as a multi-threaded entity. Apart from increasing the framework's modularity, this scheme allows independent components to work concurrently and therefore decrease the overall delay when handling MA objects, especially when the rate of incoming MAs is relatively high or bursty. The main building blocks of a MAS server are illustrated in Figure 4 with their implementation detailed in the following subsections.

**Mobile Agent Listener (MAL)**

The MAL is a daemon, which listens for incoming MAs on a well-known port. Upon the arrival of an MA, its state is decompressed and de-serialized. The MA is then authenticated by the Security Component (SC). If the authentication is successful (the MA has been created by a trusted host), the MA is first registered by the Mobile Agent Register (MAR) component. A separate thread is subsequently created for the MA's execution and assigned a certain priority level which equals a 'priority value' attribute carried by the MA. That way, critical management tasks may be prioritized against other pending executing threads on the local device, ensuring their timely execution.

The MAL then binds the MA object to the local MAS server to enable the interaction of the two parties (allow the MA to invoke MAS methods and vice-versa). Last, the MA's execution is started.

**Security Component (SC)**

This component acts as the system's protective barrier. First, the SC verifies the authenticity of the received MA through the use of security keys, ensuring that only trusted agents (dispatched by authorized hosts) are instantiated. In addition, it authorizes the actions performed by the locally executing MAs and ensures the privacy of sensitive management data returned to the manager station. The RSA (Rivest-Shamir-Adleman) algorithm [12] has been implemented providing both *authentication* and *encryption* features. It is noted that data encryption is performed only if requested by the administrator at the MA's creation time.

The security of MAS entities has been strengthened by introducing *authorization* features that restrict the authority domain of visiting MAs upon legacy systems. Specifically, the JVM's standard *Security Manager* (SM) has been extended to prevent MAs from directly accessing files, creating sub-processes, shutting down the MAS application, etc.

**Mobile Agent Register (MAR)**

MAS entities keep active control of the MA objects executing on their local devices. In particular, the MAR component maintains a hashtable, using MA IDs as a primary key and including a list of "*MA_Info*" objects, each mapped to an MA object running on the local host. MA_Info objects are basically records that include a reference to the MA object they correspond to (allowing the MAS server to perform a number of actions upon it, i.e. invoke its methods) and additional information related to the MA, such as the MA's class name, the execution frequency of its management task, its arrival time, its execution status (activated, de-activated, suspended), etc. The hashtable is dynamically updated whenever an MA arrives or leaves the host.

The information maintained by the MAR can be remotely obtained by the manager application through a Remote Method Invocation (RMI) call and presented in the Graphical User Interface (GUI) (see Figure 5). The administrator may then select any of the MA instances and perform a number of actions upon them, e.g. suspend, resume or activate an MA, modify the management task's polling frequency, etc.

---

[1] The installation of a MAS server on every managed device since SNMP interactions between MAs hosted by a MAS server and remote SNMP agents are also possible

Figure 5. On-line visual profiling of MA objects executing on a network device

**Service Facilitator (SF)**

Upon successful authentication, the MA is activated and provided a handle to the Service Facilitator (SF) component, which serves as an interface between MAs and services offered to them. SF generally includes the 'know-how' of the services offered to incoming MAs, i.e. all the functionality needed by the MA objects to perform their decentralized management tasks.

At its current implementation, the SF component is solely oriented to NSM applications; as such, it basically offers a library of methods that facilitate the interaction of MA objects with the local SNMP agent. In a typical scenario, an MA passes an arbitrary number of object MIB OID strings to an SF method, which performs the SNMP query and returns the requested values. These values are subsequently processed, if necessary, by the MA. The value acquired, either directly by the system or as a result of computation, is passed to the SC sub-system, encrypted and stored within the MA's data folder.

**Migration Facility Component (MFC)**

The role of the MFC is to dispatch upon request an MA object to a specific network device. An MA transfer may be requested through calling the move() method of the MFC. The MFC component is part of both the MAS servers and the manager application. In principle, its operation is the inverse of the MAL's operation. Namely, it establishes a network connection, compresses and serializes the MA state, with the resulted byte stream directed into the connection's output stream. Following the MA's migration, the MFC deletes the MA_Info object associated with to the dispatched MA by invoking the delete() method of the MAR component.

**Mobile Agent Server RMI Server**

The RMI server is implemented as a separate thread, which can optionally be instantiated by the main MAS thread (this is indicated by the user in the command line that starts the MAS application). The role of the RMI server is to enable remote interaction of the manager with distributed MAS entities and, hence, allow the administrator to obtain reports regarding the CPU and memory load profile of network devices or the number and type of MA objects currently executing on them. It also allows the administrator to perform a number of actions upon the MAs (see Figure 5).

**Resource Inspection Application (RIA)**

The RIA is an application developed in C programming language, which runs outside the boundary of the MAS server. Its purpose is to monitor the usage of local resources in terms of CPU and memory load. RIA is linked to the MAS application via the Java Native Interface (JNI) that enables the inter-operation of Java applications with programs written in other languages. The motivation behind RIA's development has been to allow the administrator and incoming MAs to obtain information regarding network devices load.

**Network Discovery Daemon (NDD)**

The NDD is a thread instantiated by the main MAS thread, whose purpose is to discover the station where the manager application runs, while also making the local host 'visible' to the manager. Its operation is similar to the Network Discovery thread of the manager application.

**Class Loader Daemon (CLD)**

The CLD is a daemon controlled by the MAS thread, whose role is to wait for MA class definitions sent by the manager application at the time that a new management service is introduced.

In a typical MAP, when an MA object arrives at a host, its bytecode is retrieved from the network stream and stored in the default JVM's Class Loader (CL) cache. This CL works fine as long as MAs' definitions remain unchanged; it is not able though to distinguish different versions of the same class, due to a limitation of the default JVM's CL [18]. That is, the older (cached) version of a class is still loaded even if an updated version has been uploaded. This represents a major problem for MA-based management applications wherein management tasks are often modified at runtime and expected to take immediate effect.

To the best of our knowledge, none of the existing MAPs copes with this MAs *'versioning'* problem and, as a result, the administrator is forced to 'reboot' all the MAS servers where the old version of the updated MA class is already stored to allow modifications to take effect. This solution is certainly not an adequate solution for networks comprising hundreds of managed nodes. To this end, we have extended the standard JVM implementing a custom CL, the Mobile Agent Class Loader (MACL), which is able to identify different versions of the same MA class and enable the execution of the most recent version while disposing the old one.

*D. Mobile Agent Generator (MAG)*

The MAG, is a tool for constructing customized, service-oriented MAs. In the context of this article, generated MAs are designed to poll static management (SNMP) agents according to certain operational function

requirements. A GUI, dedicated to the MAG tool, allows the operator to:

- assign a name to the MA;
- specify the generic type of service this MA is intended to carry out;
- define the MA's functional requirements, i.e. determine its operational behaviour;
- set the polling frequency to determine how often instances of the constructed MA will be launched;
- determine whether the data collected by the MA are to be encrypted and the MA itself authenticated;
- specify the class of network devices to be polled;

The MAG uses a skeleton Java source code file with empty slots filled with the user-specified MA's properties. The Java code created is then programmatically compiled through an invocation of the `compile()` method of `sun.tools.javac.Main` class. The generated Java bytecode is subsequently compressed and transferred (multicasted) through TCP connections to all operating agent hosts (see Figure 6). On the agent side, the CLD receives and decompresses the transmitted bytecode, validates the included Java class and stores it in a designated space.

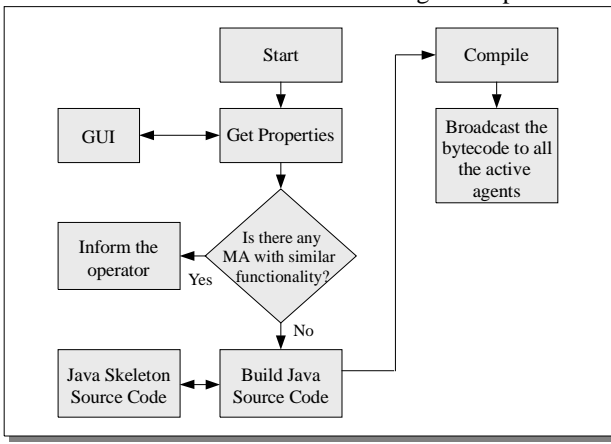

Figure 6: Mobile Agents Generator functional diagram

It should be emphasized that the transfer of the MA bytecode is performed only *once*, at the MA construction time. From that point onwards the transfer of persistent state, obtained from serializing the MA instance, is sufficient for MAS entities to recognize the incoming MA and recover its state. In contrast, most available MAPs apply a policy that requires the transfer of both the MA's bytecode and state, resulting in higher demand on network resources. To illustrate, in our implementation the ratio (bytecode size):(state size) typically lies in the range 10:1 to 15:1.

The functionality of MAs created by the MAG may only be an extension of limited generic service types. These types are designed as sub-classes of the MA 'super-class', specifying general patterns of MA-based NSM tasks. MAs constructed by the MAG tool extend one of these sub-classes, refining their functionality and defining service-specialized MAs.

The use of the MAG tool brings forth a number of advantages:

- The MAG ensures that the framework remains sufficiently flexible by enabling on-the-fly construction of service-oriented agents, without the need for reconfiguration, re-installation or re-instantiation of either the manager or the agent applications. The MAG functionality can easily be extended so as to cover a wider range of management tasks.
- Ease in introducing new management tasks in an automated manner, transparent to the user. The productivity of the development process is increased through reducing the time needed to develop an MA and improving reliability as a result of code reusability.
- The network administrator does not need to be aware of the implementation details behind the introduced management service or the management framework, nor to have programming experience.

## III. CONCLUSIONS & FUTURE WORK

MAPs tailored to NSM applications should not only provide an programmable environment for performing decentralized management tasks but also to satisfy a number of design requirements in terms of flexibility, functionality, security and demand on systems and network resources.

In this paper, we introduced a complete MAP research prototype that sufficiently addresses all the aforementioned concerns. Among others, our MAP incorporates the following novel features: (a) a lightweight code distribution scheme, (b) a class loading mechanism that allows the modification of MA-based NSM tasks at runtime, (c) a tool that supports the user-friendly customization of service-oriented MAs, (d) a component that builds near-optimal network-dependent MA itineraries. In addition, it satisfies other important NSM-related requirements, such as lightweight footprint on systems resources, security (authorization, authentication and encryption), fault tolerance, modularity, platform-independence, etc. Our MAP's performance has been evaluated in realistic management application scenarios.

Our future research will include a performance comparison in terms of latency and network overhead between our proposed MAP and general-purpose MAPs, e.g. [1][2] (unfortunately, management-oriented MAPs are not publicly available).